\documentclass[11pt,a4paper,reqno]{amsart} 
\usepackage{amsthm,amsmath,amsfonts,amssymb,amsxtra,appendix,bookmark,dsfont,bm,mathrsfs,amstext,amsopn,mathrsfs,mathtools,comment,cite,hyperref,color}
\usepackage{tikz}\usetikzlibrary{arrows}
\usepackage{makecell} 
\usepackage{stmaryrd} 
\usepackage{algorithm}
\usepackage{ifthen}\newboolean{versionlongue}\setboolean{versionlongue}{false} 

\newtheorem{theorem}{Theorem}\newtheorem{remark}{Remark}\newtheorem{lemma}{Lemma}

\newcommand\cP{\mathcal{P}} \newcommand\cH{\mathcal{H}}        \newcommand\cE{\mathcal{E}} \newcommand\cL{\mathcal{L}}    \newcommand\cS{\mathcal{S}}   \newcommand\cV{\mathcal{V}} \newcommand\cI{\mathcal{I}}\newcommand\cN{\mathcal{N}}\newcommand\cJ{\mathcal{J}}

\let\C\relax\newcommand{\C}{\mathbb{C}}\newcommand{\Z}{\mathbb{Z}}\newcommand{\R}{\mathbb{R}}\newcommand{\N}{\mathbb{N}} %
\newcommand{\bbH}{\mathbb{H}}\newcommand{\bbK}{\mathbb{K}}
\newcommand{\Np}{\mathbb{N} }
\newcommand{\Nz}{\mathbb{N} \cup \acs{0}}

\DeclareMathOperator{\tr}{Tr}
\def\d{{\rm d}}
\newcommand{\parmi}[2]{\begin{pmatrix} #2 \\ #1 \end{pmatrix}}

\newcommand{\floor}[1]{\left\lfloor #1 \right\rfloor}
\newcommand{\roof}[1]{\left\lceil #1 \right\rceil}
\DeclarePairedDelimiter\ceil{\lceil}{\rceil}

\newcommand{\nor}[2]{ \left\| #1 \right\|_{#2} } 

\newcommand{\ep}{\varepsilon} 
\let\p\relax\newcommand{\p}{\Psi} 
\newcommand{\na}{\nabla} 
\newcommand{\ro}{\rho}

\newcommand{\f}[2]{\frac{#1}{#2}}
\newcommand{\bul}{$\bullet$ \hspace{0.1cm}}
\newcommand{\mymax}[1]{\underset{\substack{#1}}{\text{\normalfont{max}}}\hs}

\newcommand{\myinf}[1]{\underset{\substack{#1}}{\text{\normalfont{inf}}}\hs}

\newcommand{\ssu}{\begingroup\textstyle\sum\endgroup} 

\newcommand{\ind}[1]{_{\textup{#1}}} 
\renewcommand{\ge}{\geqslant}
\renewcommand{\le}{\leqslant}

\newcommand{\tx}[1]{\text{#1}}
\newcommand{\df}{:=}
\newcommand{\bst}{\hspace{0.1cm} \Big| \hspace{0.1cm}} 
\newcommand{\st}{\hspace{0.1cm} \bigr\vert \hspace{0.1cm}} 
\newcommand{\sch}{\mathfrak{S}} 
\newcommand{\ii}{\infty}
\newcommand{\hs}{\hspace{0.1cm}}
\newcommand{\bhs}{\hspace{1cm}}

\newcommand{\ud}{\frac{1}{2}}

\newcommand{\vl}[1]{ \ifthenelse{\boolean{versionlongue}}{#1}{} }

\newcommand{\kincan}{{\ssu_{i=1}^N (-\Delta_i)}}

\newcommand{\pa}[1]{\left( #1 \right)} 
\newcommand{\bpa}[1]{\big( #1 \big)} 
\newcommand{\bbpa}[1]{\Big( #1 \Big)} 
\newcommand{\acs}[1]{\left\{ #1 \right\}} 
\newcommand{\seg}[1]{\left[ #1 \right]} 
\newcommand{\ab}[1]{\left|#1\right|} 
\newcommand{\ps}[1]{\left< #1 \right>} 

\newcommand{\Nm}{\overline{N}}
\newcommand{\opeind}{\circ}
\newcommand{\aea}{\beta}
\newcommand{\ope}{\circ_\aea}
\newcommand{\F}{F}
\newcommand{\App}{\Theta}

\newcommand{\dl}[1]{\delta_{\aea,\textup{#1}}}

\title[Representability of density-entropy pairs]{Mixed state representability\\of entropy-density pairs}
\author[L. Garrigue]{Louis Garrigue}
\address{Laboratoire ``analyse géométrie modélisation'', CY Cergy Paris Université, 95302 Cergy-Pontoise, France}
\email{louis.garrigue@cyu.fr}

\begin{document} 
\maketitle
\begin{abstract}
We show the representability of density-entropy pairs with canonical and grand-canonical states, and we provide bounds on the kinetic energy of the representing states.
\end{abstract}
\date{\today}

\section{Introduction} 

One of the basic questions arising in the variational approach of many-body quantum mechanics, and raised by Coleman \cite{Coleman63}, is whether some quantities like one-body or two-body density matrices, or other reduced quantities, are the ones of an $N$-body fermionic wavefunction having finite kinetic energy. For instance the ground state of a system can be extracted from the minimization of the energy over the set of two-particle density matrices coming from fermionic wavefunctions \cite{Lowdin55a,Coleman63,ColYuk00}, and knowing a simple characterization of this set would enable to reduce the exact many-body problem to a low-dimensional one. Those issues are called the pure-state representability problems. For one-body densities it was solved by Harriman in \cite{Harriman81}, using determinantal functions, and was later put on rigorous grounds by Lieb in \cite[Theorem~1.2]{Lieb83b}. Then the problem of pure-state representability of current densities was addressed by Lieb and Schrader in \cite{LieSch13}, solved with explicit bounds when the velocity field is curl-free. The case of pure-state representability of magnetizations was solved by Gontier in \cite{Gontier13,Gontier16}, also with explicit bounds. Finally, see \cite{Coleman63} and \cite[Theorem 3.2]{LieSei09} for a solution of the mixed state representability of one-body density matrices. On the contrary, representing pair densities \cite{Pistol04,AyeDav06,Pistol06} and two-body (or $q$-body) density matrices \cite{Coleman63,Kummer67,Smith66,Mazziotti12} with $N$-body fermionic states are important but still open problems, simple necessary and sufficient conditions are not known. 

In this document, we address the corresponding problem of state representability in the quantum thermodynamic ensembles. More precisely, we build mixed states having prescribed density and entropy $(\ro,S)$ in the canonical case, and having prescribed density, mean number of particles and entropy $(p,\Nm,S)$ in the grand-canonical case, for both bosons and fermions, providing explicit bounds on the kinetic energy of those mixed states. Finally, we can apply those results to define internal Levy-Lieb functionals in the Thermal DFT \cite{PriPitGro14} configuration.

\section{Main result}

\subsection{Definitions}
We use $\N = \{1,2,3,\dots\}$ as the set of non-vanishing integers. For any Hilbert space $\cH$, we denote by $\cL\pa{\cH}$ the set of bounded operators of $\cH$. The first Schatten space is defined as the set $\sch_1(\cH) = \acs{\Gamma \in \cL\pa{\cH}, \tr \ab{\Gamma} < +\infty}$ endowed with the trace norm, and it is a Banach space \cite{Simon79}. 

\subsubsection{States}%
\label{ssub:States}

For any $N\in \N$, the set of $N$-body canonical states (depending on if we consider fermions or bosons) is
 \begin{multline*}
\cS^N\ind{can} \df \big\{\Gamma_N = \Gamma_N^* \ge 0 \st \Gamma_N \in \sch_1 \bpa{L\ind{a/s}^2\bpa{\R^{dN}}}, \\
	\tr \Gamma_N = 1, \tr \kincan \Gamma_N < +\infty \big\},
 \end{multline*}
 where $L^2\ind{a}(\R^{dN}) \df \wedge^N L^2(\R^d)$ is the set of fermionic $N$-body states (``a'' stands for antisymmetric and $\wedge$ denotes the antisymmetric product), and $L^2\ind{s}(\R^{dN}) \df \vee^N L^2(\R^d)$ is the set of bosonic $N$-body state (``s'' stands for symmetric and $\vee$ denotes the symmetric product, see~\cite[Section 1.1]{Lewin11} for instance for a precise definition). Moreover, the inequality $\Gamma \ge 0$ is in the sense of quadratic forms. The set of grand-canonical states is
 \begin{align*}
	 \cS\ind{gc} \df \acs{\Gamma = \Gamma^* \ge 0 \st \Gamma \in \sch_1 \bpa{ \oplus_{n=0}^{+\ii} L\ind{a/s}^2\bpa{\R^{dn}}}, \tr \Gamma = 1, \tr \bbK \Gamma < +\infty}
 \end{align*}
 where $\oplus_{n=0}^{+\ii} L\ind{a/s}^2\bpa{\R^{dn}}$ is the fermionic/bosonic Fock space, and where
 \begin{align*}
 \bbK := \bigoplus_{n=0}^{+\infty} \; \sum_{j=1}^n \pa{-\Delta_j}
 \end{align*}
is the grand-canonical kinetic energy operator. We have $\cS^N\ind{can} \subset \cS\ind{gc}$ for any $N \in \N$.

\subsubsection{Densities}%
\label{ssub:Densities}

Also denoting by $\Gamma_N$ the kernel of an $N$-body mixed state $\Gamma_N \in \cS^N\ind{can}$, we define the one-body density of a canonical state $\Gamma_N$ by
 \begin{align*}
 \ro_{\Gamma_N}(x) := N \int_{\R^{d(N-1)}} \Gamma_N(x,x_2,\dots,x_N;x,x_2,\dots,x_N) \d x_2 \cdots \d x_N.
 \end{align*}
The density $\ro_\Gamma$ of a grand-canonical state $\Gamma = \oplus_{n=0}^{+\infty} \Gamma_n$, and its normalized density $p_\Gamma$ are defined by
 \begin{align*}
 \ro_\Gamma := \sum_{n=1}^{+\infty} \ro_{\Gamma_n}, \qquad\qquad  p_\Gamma := \f{\ro_\Gamma}{\int_{\R^d} \ro_\Gamma}.
 \end{align*}

A classical construction from Harriman and Lieb \cite{Harriman81,Lieb83b} shows that a density $\ro \in L^1(\R^d,\R_+)$ is the one-body density of a physical pure wavefunction $\p \in  H^1_{\tx{a}}(\R^{dN}) := \pa{L^2\ind{a}\cap H^1}(\R^{dN}) , \int_{\R^{dN}} \ab{\p}^2 = 1$ if and only if $\ro \in \cI^N$, where
 \begin{align*}
	 \cI^N \df \acs{\ro \in L^1(\R^d,\R) \bst \ro \ge 0, \sqrt{\ro} \in H^1(\R^d), \int_{\R^d} \ro = N }.
 \end{align*}
 Moreover, there is an explicit kinetic energy bound for $\p$,
 \begin{align*}
	 \int_{\R^{dN}} \ab{\na \p}^2 \le (4\pi)^2 N^2 \int_{\R^d} \ab{\na \sqrt{\ro}}^2.
 \end{align*}

 \subsubsection{Density-entropy pairs}%
 \label{ssub:Density-entropy pairs}
 
 For $\Gamma \in \cS\ind{gc}$, we consider the standard entropy $S_\Gamma \df -\tr \Gamma \ln \Gamma$, where the trace is taken on the natural operator Hilbert space on which $\Gamma$ is defined. In Thermal DFT \cite{PriPitGro14}, which is an approach of quantum thermodynamics, the corresponding representability problem consists in characterizing the set of pairs $(\ro,S) \in L^1(\R^d) \times \R_+$ coming from mixed states. In this case, the set of admissible density-entropy pairs is defined by
 \begin{multline*}
	 \cI^N\ind{can}  \df \{ (\ro,S) \in L^1(\R^d) \times \R_+ \st \exists \Gamma_N \in \cS^N\ind{can}, \ro_{\Gamma_N} = \ro, \\
	 S_{\Gamma_N} = S, \tr \ssu_{i=1}^N (-\Delta_i) \Gamma_N < + \ii \},
 \end{multline*}
 for the canonical setting, and
 \begin{multline*}
	 \cI\ind{gc}  \df \{ (p,\Nm,S) \in L^1(\R^d) \times ]0,+\infty[ \times \R_+ \st \exists \Gamma \in \cS\ind{gc},  \\
	 p_{\Gamma} = p,\tr \cN \Gamma = \Nm, S_\Gamma = S, \tr \bbK \Gamma < + \ii \},
 \end{multline*}
 for the grand-canonical setting, where $\cN = \oplus_{n=0}^{+\ii} \; n$ is the particle number operator. The mean number of particles is
\begin{align*}
\Nm_\Gamma := \tr \cN \Gamma = \int_{\R^d} \ro_\Gamma.
\end{align*}

\subsubsection{Hamiltonian}%
\label{ssub:Hamiltonian}

Finally we define the $N$-body Hamiltonian
 \begin{align*}
	 H^N(v) \df \sum_{i=1}^N (-\Delta_i + v(x_i)) + \sum_{1 \le i < j \le N} w(x_i-x_j),
 \end{align*}
 acting on canonical states of the $N$-particle sector $\otimes^N L^2(\R^{d})$, and 
\begin{align*}
 \bbH(v) \df \bigoplus_{n=0}^{+\ii} \; H^n(v)
\end{align*}
acting on the Fock space. The function $v$ is the electric potential while $w$ is the interaction potential. We will always assume that $v, w \in L^{q}\ind{loc}(\R^d)$, $w \ge 0$, where
 \begin{align}\label{exps}
	 q = 1 \tx{ for } d=1,\bhs q>1 \tx{ for } d=2, \bhs q = \f d2 \tx{ for } d \ge 3. 
 \end{align}

\subsection{Representability of density-entropy pairs}
\label{sub:rep_thms}

\begin{theorem}[Representability of $(\ro,S)$ with canonical states]\label{thm:rep_can_S}
Take $\ro \in \cI^N$ and $S \in \R_+$. There exists $\Gamma_N \in \cS^N\ind{can}$ such that $\ro_{\Gamma_N} = \ro$, $S_{\Gamma_N} = S$, and
\begin{align}\label{ineq:thm1} 
\tr \ssu_{i=1}^N (-\Delta_i) \Gamma_N \le \pa{1+ 4\pi^2 N^2 \pa{\pa{e^S +2}^{\f 1N} - \chi +\tfrac 4N}^2} \int_{\R^d} \ab{\na \sqrt{\ro}}^2.
\end{align}
where $\chi = 1$ for bosons and $\chi = 0$ for fermions. We have thus
\begin{align}\label{eq:cI_can}
\cI^N\ind{can} = \cI^N \times \R_+.
\end{align}
\end{theorem}
A proof is given in Section~\ref{sec:proof_thm_1}. In~\eqref{ineq:thm1}, the bosonic and fermionic cases can be different, for instance if we let $N \rightarrow +\infty$ and keep $\ro$ and $S$ fixed, the right hand side of~\eqref{ineq:thm1} is bounded in $N$ for bosons while it diverges to $+\infty$ proportionally to $N^2$ for fermions.

We then treat grand-canonical systems, in the case where the mean number of particles is not controlled.
\begin{theorem}[Representability of $(p,S)$ with grand-canonical states]\label{thm:rep_gc_S}
Take $p \in \cI^1$ and $S \in \R_+$. There exists $\Gamma \in \cS\ind{gc}$ such that $p_\Gamma = p$, $S_\Gamma = S$ and
\begin{align*}
\tr\bbK\Gamma \le 5\times 2^{10}  (S+4)^2 \int_{\R^d} \ab{\na \sqrt{p}}^2, \bhs \Nm_{\Gamma} \le 2^4(S+4),
 \end{align*}
both for fermions and bosons.
\end{theorem}
A proof is provided in Section~\ref{sec:proof_thm2}.

\subsection{Representability of density, mean number of particles, and entropy}%
\label{sub:Representability of density, mean number of particles, and entropy}

Finally, we treat the grand-canonical case where we control $S$, $p$ and $\Nm$ at the same time. Let us define the entropic function
\begin{align}\label{eq:def_s(x)} 
s(x) := -x\ln x - \pa{1-x} \ln \pa{1-x},
\end{align}
on $[0,\frac 12]$, which is strictly increasing and spans $[0,\ln 2]$. The reciprocal function of $s$ is denoted by $s^{-1} : [0,\ln 2] \rightarrow [0,\f{1}{2} ]$, it is strictly increasing, and for any $S \in [0,+\infty)$, we define
\begin{align}\label{def:xiS}
\xi(S) := \left\{
\begin{aligned}
& s^{-1}(S) & \text{ if } S \le \ln 2, \\
& 1 & \text{ otherwise}.
\end{aligned}
\right.
\end{align}

\begin{theorem}[Representability of $(p,\Nm,S)$]\label{thm:rep_gc_N_S}
Take $p \in \cI^1$, $S, \Nm \in \R_+ \backslash \{0\}$. There exists $\Gamma \in \cS\ind{gc}$ such that $(p_\Gamma,\Nm_\Gamma,S_\Gamma) = (p,\Nm,S)$ and having kinetic energy bounded by 
\begin{align}\label{ineq:kin}
\tr \bbK \Gamma \le 
5\pi^2 \pa{ \f{\Nm}{\xi(S)} +1}^3 \pa{e^{\f{S}{\Nm} \pa{ 1 + \f{\xi(S)}{\Nm}}} + 6}^2 \int_{\R^d} \ab{\na \sqrt{p}}^2,
\end{align}
both for fermions and bosons, where $\xi(S)$ is defined in \eqref{def:xiS}. Moreover,
\begin{align}\label{eq:cIgc}
	\cI\ind{gc} =\cI^1  \times \bbpa{\bpa{ (0,+\infty) \times (0,+\infty)} \cup \pa{\N \times \{0\}}}
\end{align}
\end{theorem}
A proof is provided in Section~\ref{sec:proof_thm_3_rep_gc_N_S}. More explicitely,~\eqref{eq:cIgc} tells that a $(p,\Nm,S)$ is representable if $\Nm \in \N$ or if $S \neq 0$. Indeed, it is not possible to build a state $\Gamma$ with $\Nm_\Gamma \notin \N$ and $S_\Gamma = 0$. 

The following lemma provides a better idea of the behavior of $s^{-1}$ in a neighborhood of $0^+$. 
\begin{lemma}[Behavior of $s^{-1}$ close to $0^+$]\label{lem:behavior_s_inv} 
For any $\ep \in [0,\f 12]$, there exists $y_\ep \in (0,\ln 2)$ such that for any $y \in [0,y_\ep]$,
\begin{align}\label{eq:bound_s_inv} 
y^{1+\ep} \le s^{-1}(y) \le y.
\end{align}
\end{lemma}
A proof of~\eqref{eq:bound_s_inv} is provided in Section~\ref{sub:Proof of bound_s_inv}. Let us make a few remarks on Theorem~\ref{thm:rep_gc_N_S}. 

\begin{remark} Concerning the right hand side of~\eqref{ineq:kin},

\bul when $\Nm \rightarrow 0^+$, it diverges proportionally to $e^{\f{2S}{\Nm} \pa{ 1 + \f{\xi(S)}{\Nm}}}$,

\bul when $S \rightarrow 0^+$, it diverges proportionally to $\pa{s^{-1}(S)}^{-3}$, which diverges more slowly than $S^{-3(1+\ep)}$ for any $\ep > 0$ by Lemma~\ref{lem:behavior_s_inv},

\bul when $\Nm \rightarrow +\infty$, it diverges proportionally to $\Nm^3$,

\bul when $S \rightarrow +\infty$, it diverges proportionally to $e^{\f{2S}{\Nm}\pa{ 1+ \f{1}{\Nm} }}$.
\end{remark}

\begin{remark}
In all the results, the condition $\sqrt{\rho} \in H^1(\R^d)$ cannot be dropped because of the Hoffmann-Ostenhof inequality~\cite{Hof77} (see also~\cite[(2.11)]{Lewin11}) being $\int_{\R^d} \ab{\na \sqrt{\rho_\Gamma}}^2  \le \tr \bbK \Gamma$ for any $\Gamma \in \cS\ind{gc}$.
\end{remark}

\section{Definition of internal energies}%
\label{sec:Definition of internal energies}
In this section, we present an application of the representability results of Section~\ref{sub:rep_thms}. More precisely, we show that they enable us to define internal energies.

\subsection{Free energies} 
To prepare, we define the free energy functionals, for $v \in \pa{L^q + L^{\infty}}(\R^d)$ and $T \in \R_+$,
\begin{align*}
\cE^{N}_{v,T}(\Gamma) := \tr H^N(v) \Gamma - TS_\Gamma,\qquad \cE_{v,T}(\Gamma) := \tr \bbH(v) \Gamma - TS_\Gamma,
\end{align*}
and the free energies
\begin{align*}
E^N\ind{can}(v,T) \df \myinf{\Gamma \in \cS^N\ind{can}} \cE^{N}_{v,T}(\Gamma), \bhs E\ind{gc}(v,T,\mu) \df \myinf{\Gamma \in \cS\ind{gc}} \cE_{v-\mu,T}(\Gamma).
\end{align*}
First, $E^N\ind{can}(v,T)$ is jointly concave in $(v,T)$ and $E\ind{gc}(v,T,\mu)$ is jointly concave in $(v,T,\mu)$. They are increasing in $v$ and decreasing in $T$.

\subsection{Free Levy-Lieb functionals}

For $\ro \in \cI^N$, $p \in \cI^1$, $\Nm \in \R_+$, the thermal Levy-Lieb functionals are
\begin{align}\label{fre}
	F^{N,T}\ind{can}(\ro) \df \hspace{-0.2cm}\myinf{\Gamma \in \cS^{N}\ind{can} \\ \ro_{\Gamma} = \ro} \hspace{-0.2cm} \pa{ \tr H^N(0) \Gamma - TS_\Gamma}, \hs\hs
F^{T}\ind{gc}(\Nm p) & \df  \hspace{-0.2cm}\myinf{\Gamma \in \cS\ind{gc} \\ \ro_{\Gamma} = \Nm p} \hspace{-0.2cm}\pa{ \tr \bbH(0) \Gamma - TS_\Gamma},
\end{align}
which enable to recover the free energies by Legendre transforms. On $\cI^N$, we have $F^{N,T}\ind{can} = F^T\ind{gc}$. Some properties of those Levy-Lieb functionals at zero temperature were recently studied in \cite{LewLieSei19,LewLieSei22}.

\subsection{Internal Levy-Lieb functionals}
There is a second way of defining Levy-Lieb functionals, in terms of internal quantities only, that is no external field is involved. Their definition is an application of our representability theorems.

For $(\ro,S) \in \cI^N \times \R_+$, and $(p,S,\Nm) \in \cI\ind{gc}$ in the grand-canonical case, we define 
\begin{equation}\label{ll}
	F^N\ind{can}(\ro,S) \df \hspace{-0.3cm} \myinf{\Gamma \in \cS^{N}\ind{can} \\ \pa{\ro_{\Gamma}, S_\Gamma} = (\ro,S)} \hspace{-0.3cm}\tr H^{N}(0) \Gamma, \bhs F\ind{gc}(\Nm p,S) \df \hspace{-0.3cm}\myinf{\Gamma \in \cS\ind{gc} \\ \pa{\ro_{\Gamma}, S_\Gamma} = (\Nm p ,S)} \hspace{-0.3cm}\tr \bbH(0) \Gamma,
\end{equation}
and call them internal Levy-Lieb functionals. They are finite by the representability results Theorem~\ref{thm:rep_can_S} and Theorem~\ref{thm:rep_gc_N_S}. The internal Levy-Lieb functionals \eqref{ll} enable to recover the free energies by Legendre transforms,
 \begin{align*}
	 E^N\ind{can}(v,T)  & = \myinf{(\ro,S)   \in \cI^N \times \R_+} \pa{F^N\ind{can}(\ro,S) + \int_{\R^d} \ro v - T S}, \\
	 E\ind{gc}(v,T,\mu) & = \myinf{(p,S,\Nm) \in \cI\ind{gc}} \pa{F\ind{gc}(\Nm p,S) + \Nm \int_{\R^d}  p (v-\mu) -  T S }.
 \end{align*}
We remark that we could also define Lieb functionals, similarly as in \cite[(3.17)]{Lieb83b}. They are convex and weakly lower-semicontinuous, and a reconstruction formula similar as \cite[(3.20)]{Lieb83b} also holds.

We recall that the map $(v,T) \mapsto (\ro_{\Gamma},S_{\Gamma})$, where $\Gamma$ is the Gibbs state of $(v,T)$, is injective. This is the Hohenberg-Kohn theorem for systems at positive temperature, see \cite{Mermin65} and \cite[Theorem~4.1]{Garrigue19b}, and this exposes the $(v,T) - (\ro,S)$ duality. In other words, the information of the internal quantities $(\ro,S)$ enables to recover the information of the external fields $(v,T)$.
o

\section{Preliminary constructions for the proofs}%
\label{sec:Preliminary constructions for the proofs}

In this first section, we present of two sets of functions, denoted by $\phi_k$ and $\p_k^\aea$, which are relevant for the proofs of all our results.

We define the kinetic energy $T(\cdot)$ of pure states $\p$, of canonical states $\Gamma_N$, and of grand-canonical states $\Gamma$ as
 \begin{align*}
	 T(\p) := \int_{\R^{dN}} \ab{\na \p}^2, \quad\quad T(\Gamma_N) := \tr \kincan \Gamma_N, \quad\quad T(\Gamma) := \tr\bbK\Gamma.
 \end{align*}
For $p \in \Nz$ and $q \in \Z$, we use the notation
 \begin{align}\label{def:parmi}
	 \parmi{p}{q} = \f{q!}{p!(q-p)!} \tx{ when } p \le q, \bhs \parmi pq = 0 \tx{ when } q < p \tx{ or } q < 0.
 \end{align}

\subsection{Harriman-Lieb orbitals}%
\label{sub:Harriman-Lieb orbitals}

 We use the construction of Harriman and Lieb \cite{Harriman81,Lieb83b}, which corresponds to the pure case $S=0$. Let us take $\rho \in \cI^N$. We recall Harriman's function 
\begin{align*}
f(x^1) \df \f{2\pi}{N} \int_{-\ii}^{x^1} \d s \int_{\R^{d-1}} \d x^2 \cdots \d x^d \ro(s,x^2, \dots,x^d),
\end{align*}
where $x = (x^1, \dots, x^d) \in \R^d$. The one-particle orbitals $\phi_k : \R^d \rightarrow \C$ at stake are
\begin{align}\label{deforb}
\phi_k(x) \df \sqrt{\f{\ro(x)}{N}}e^{ikf(x^1)},
\end{align}
for $k \in \Z$. They form an orthonormal family because $\ps{\phi_\ell,\phi_m} = \delta_{\ell-m}$, they have density $\ab{\phi_k}^2 = \ro/N$, and we will say that $k$ is the momentum of $\phi_k$. Then it is proved in \cite[Proof of Theorem 1.2]{Lieb83b} that
\begin{align}\label{eq:kin_rep}
	\int_{\R^d} \ab{\na \phi_k}^2 \le \f 1{N} \pa{1+ 16\pi^2 k^2} \int_{\R^d} \ab{\na \sqrt{\ro}}^2.
\end{align}
In all this document, $\phi_k$ will denote those functions.

\subsection{Constraints on ordered sets $\bpa{\Psi^\aea_k}_{k \in \N}$ of pure states}%
\label{sub:Definition of ordered sets}

\subsubsection{Definition}%
\label{ssub:Definition_Psi}

Take $\rho \in \cI^N$, we define
\begin{align*}
\cJ^{\text{a}}_N &:= \acs{\pa{j_i}_{1 \le i \le N} \;|\;j_1, \dots, j_N \in \Z, j_1 < \dots < j_N}, \\
\cJ^{\text{s}}_N &:= \acs{\pa{j_i}_{1 \le i \le N} \;|\;j_1, \dots, j_N \in \Z, j_1 \le \dots \le j_N},
\end{align*}
where ``a'' stands for ``antisymmetric'' (fermionic case) and ``s'' for ``symmetric'' (bosonic case). We define the operators 
\begin{align*}
\opeind_{\text{a}} := \wedge, \qquad \qquad \opeind_{\text{s}} := \vee.
\end{align*}
and for any $\aea \in \{\text{a},\text{s}\}$ and  any $J = (j_1,\dots,j_N) \in \cJ^{\aea}_N$, the many-body pure states
\begin{align}\label{eq:notation_PhiJ}
\Phi_{J}^\aea := \pa{\ope}_{a \in J} \phi_a = \phi_{j_1} \ope \cdots \ope \phi_{j_N},
\end{align}
respecting $\ro_{\Phi_J^\aea} = \ro$. For $\aea \in \{\text{a},\text{s}\}$, we define $\dl{a} = 1$ if $\aea = \text{a}$ and $\dl{a} = 0$ if $\aea = \text{s}$, we also define $\dl{s}$ similarly, that is $\dl{s} = 1$ if $\aea = \text{s}$ and $\dl{s} = 0$ if $\aea = \text{a}$. For $\aea \in \{\text{a},\text{s}\}$, $\ell \in \Nz$ and $\triangle \in \{ \le , = \}$, we define
\begin{align*}
\cJ^{N,\aea,\triangle}_{\ell} &:= \acs{J = (j_i)_{1 \le i \le N} \in \cJ_N^{\aea} \;\Big\rvert\; \mymax{1 \le i \le N} \ab{j_i} \triangle \;\ell },
\end{align*}
we remark that for any $\ell \in \Np$
\begin{align*}
\ab{\cJ^{N,\aea,=}_{\ell}} = \ab{\cJ^{N,\aea,\le}_{\ell}} - \ab{\cJ^{N,\aea,\le}_{\ell-1}}
\end{align*}
where for any set $E$, $\ab{E} $ denotes its cardinality. Moreover, we have that $\cJ^{N,\aea,\le}_0 = \cJ^{N,\aea,=}_0 = \{(0,0,\dots,0)\}$.

We consider ordered sets $\bpa{\p_k^\aea}_{k \in \N}$ such that $\p^\aea_1 := \Phi^\aea_{(0,\dots,0)}$, and for any $\ell \in \N$,
\begin{align}\label{eq:last_constraint_Psi}
\acs{\p_k^\aea}_{\ab{\cJ^{N,\aea,\le}_{\ell-1}} +1 \le k \le \ab{\cJ^{N,\aea,\le}_{\ell}}} = \acs{ \Phi^\aea_J \;\Big\rvert\; J \in \cJ^{N,\aea,=}_{\ell}}.
\end{align}
There are several ways to do so but we can take anyone of them, this will not change the rest of the proof. 

Some consequences of the definition are that
\begin{itemize}
\item for any $k \in \N$, there exists a unique $J \in \cJ^\aea_N$ such that $\p_k^\aea = \Phi_J^\aea$,
\item the $\p_k^\aea$'s form an orthonormal family, i.e. $\ps{\p_k^\aea,\p_\ell^\aea} = \delta_{k-\ell}$,
\item $T(\p_k^\aea)$ is not necessarily increasing in $k$,
\item $\rho_{\p_k^\aea} = \rho$.
\end{itemize}

\subsubsection{Maximum absolute value momentum}%
\label{ssub:Maximum absolute value momentum}

Once such a set $\bpa{\Psi^\aea_k}_{k \in \N}$ is chosen, given $k \in \N$, there exists $J = (j_i)_{1 \le i \le N} \in \cJ^\aea_N$ such that $\p_k^\aea= \Phi^\aea_J$, and we define
 \begin{align*}
\F^\aea_N(k) \df  \mymax{1 \le i \le N \\ \p_k^\aea= (\ope)_{q =1}^N \phi_{j_q}} \ab{j_i} = \max \bpa{\ab{j_1}, \ab{j_N}  }
\end{align*}
the maximum absolute value momentum of orbitals forming $\p_k^\aea$. From the constraint~\eqref{eq:last_constraint_Psi}, we have that
\begin{align}\label{eq:J_s}
\forall k \in \acs{\ab{\cJ^{N,\aea,\le}_{\ell-1}} +1, \dots, \ab{\cJ^{N,\aea,\le}_{\ell}}}, \qquad \F_N^\aea(k) =  \ell,
\end{align}
so $\F^\aea_N$ is an increasing map. Moreover, we can remark that $\F^\aea_N(1) = 0$ and $\F^\aea_N(2) = 1$.

\subsubsection{Bound on the kinetic energy of $\p_k^\aea$}%
\label{ssub:Bound on the kinetic energy of pk}

We now translate this into an inequality on the kinetic energy of the $\p_k^\aea$'s. Recalling that $\p_k^\aea = \Phi^\aea_J = (\ope)_{i =1}^N \phi_{j_i}$, we have 
\begin{align}\label{ineq:bound_TPsi}
	T(\p_k^\aea) &= \sum_{i=1}^N T\pa{\phi_{j_i}} \le N \mymax{1 \le i \le \F_N^\aea(k)} T(\phi_i) \nonumber \\
		& \underset{\substack{\eqref{eq:kin_rep} \\ \F^\aea_N \nearrow}}{\le} \; \pa{1+ 16\pi^2 \F_N^\aea(k)^2} \int_{\R^d} \ab{\na \sqrt{\ro}}^2.
\end{align}

\subsection{Explicit bound on $\F^\aea_N$}
 \label{sub:explicit_bound_J}

 As will be seen later in~\eqref{ineq:bound_TPsi}, we will need a bound on $\F^\aea_N(k)$. To this purpose, we first need to compute $\ab{\cJ^{N,\aea,\le}_\ell}$.
 \begin{lemma}\label{lem:bound_J}
	 For any $\ell \in \Nz$ and any $N \in \N$,
\begin{align}\label{eq:values_abs_J}
\ab{\cJ^{N,\aea,\le}_\ell} = \parmi{N}{2\ell + \dl{s} N + \dl{a}}.
\end{align}
 \end{lemma}
 \begin{proof}
For fermions, choosing an element of $\cJ^{N,\text{a},\le}_{\ell}$ consists in choosing $N$ orbitals (to build an $N$-particle determinantal state) among $2\ell +1$ available orbitals, which are $\phi_{-\ell}, \phi_{-\ell + 1},\dots, \phi_{\ell}$. Hence
\begin{align*}
\ab{\cJ^{N,\text{a},\le}_\ell} = \parmi{N}{2\ell +1}.
\end{align*}
We remark that the previous formula still holds when $N > 2\ell +1$.

For bosons, we give an instance of the first elements of an ordered set $(\p_k^\aea)_{k \in \N}$ filling the conditions of Section~\ref{ssub:Definition_Psi}. It is
 \begin{multline*}
 \p_1  \df \bigvee_{i=1}^N \phi_0, \qquad \p_2  \df \phi_1 \bigvee_{i=1}^{N-1} \phi_0, \qquad \p_3  \df \phi_{-1} \bigvee_{i=1}^{N-1} \phi_0,\\
 \p_4  \df \phi_{1} \vee \phi_{1}  \bigvee_{i=1}^{N-2} \phi_0, \qquad \p_5  \df \phi_{-1} \vee \phi_{-1}  \bigvee_{i=1}^{N-2} \phi_0, \qquad \p_6  \df \phi_{-1} \vee \phi_1  \bigvee_{i=1}^{N-2} \phi_0,
 \end{multline*}
 etc., and we go on respecting the constraints of Section~\ref{sub:Definition of ordered sets}. 

 We denote by $\zeta(m,N)$ the number of ways to fill $N$ distinguishable boxes with $m$ types of indistinguishable balls, with repetition allowed. 
For instance when the three types of balls are represented by $1$, $2$ and $3$ and when there are $3$ boxes, the configurations $221, 212, 122$ represent the same one, and the different configurations are
\begin{align*}
111, 222, 333, 123, 112, 122, 113, 133, 223, 233
\end{align*}
so $\zeta(3,3) = 10$. Computing $\zeta(m,N)$ is a well-known problem, see~\cite[Section 3.2 p6]{Riordan14} for instance, so
 \begin{align*}
\zeta(m,N) = \parmi{m-1}{N+m -1}= \parmi{N}{N+m -1}.
 \end{align*}
Choosing an element of $\cJ^{N,\text{s},\le}_\ell$ amounts to fill $N$ distinguishable boxes with $2\ell + 1$ indistinguishable objects (integer momenta from $-\ell$ to $\ell$), so
\begin{align*}
	\ab{\cJ^{N,\text{s},\le}_\ell } = \zeta(2\ell + 1,N) = \parmi{N}{N+2\ell}.
\end{align*}
 \end{proof}

We are now in a position where we can have lower and upper bounds on $\F^\aea_N$. The lower bound is similar to the upper one, the main difference is a factor $e^{-1}$. The lower bound will not be used but we derived it to show that our upper bound is satisfyingly good.

\begin{lemma}[Bounds on $\F^\aea_N$]
	For any $k, N \in \Np$, and any $\aea \in \{\text{a},\text{s}\}$,
\begin{align}\label{eq:explicit_bound_J}
\f N2 \pa{k^{\f 1N} e^{-1}- \dl{s} - \f 2N  } \le \F_N^\aea(k) \le  \f N2 \pa{k^{\f 1N}- \dl{s} + \f 4N  }.
\end{align}
\end{lemma}

\begin{proof}
First, we know that for any $k, n\in \Nz$, 
\begin{align}\label{ineq:parmi_fonda}
\pa{\f nk}^k \le \parmi{k}{n} \le \pa{\f {en}{k}}^k,
\end{align}
where formally $\pa{\f{x}{0} }^0 := 1$ for any $x \in \R_+$. For any proposition $\cP$, we define $\delta_{\cP} :=  1$ if $\cP$ is true and $\delta_{\cP} := 0$ otherwise. We have $N = 2 \floor{\f N2} + \delta_{N \in 2\N +1}$ so~\eqref{eq:values_abs_J} can be rewritten
\begin{align*}
	\ab{\cJ^{N,\aea,\le}_{\ell - \dl{s} \floor{\frac N2}}} = \parmi{N}{2\ell + \dl{a} - \dl{s} \delta_{N \in 2\N +1}}.
\end{align*}
Using~\eqref{ineq:parmi_fonda}, we have
\begin{align}\label{ineq:parmis} 
\pa{\f{2\ell -1}{N} }^N \le \ab{\cJ^{N,\aea,\le}_{\ell - \dl{s} \floor{\frac N2}}}  \le \pa{\f{e(2\ell +1)}{N} }^N
\end{align}
for any $\ell \in \N$ such that $1+\dl{s} \floor{\frac N2} \le \ell$. Since $\ab{\cJ^{N,\aea,\le}_{0} } = 1$, this also holds for any $\ell \in \Nz$ such that $\dl{s} \floor{\frac N2} \le \ell$. We defined $\F_N^\aea$ on $\N$ but we extend it to all of $\R_+$ such that it remains increasing. 

Let us find now find a lower bound for $\F^\aea_N$. We denote by $\floor{\cdot}$ the floor function and by $\ceil{\cdot} := \floor{\cdot} + 1$ the ceiling function. For any $y \in [(N+1)/2,+\infty[$,
 \begin{align*}
	 &\F^{\aea}_N\pa{ \pa{\f{2y-1}{N}}^N}  \le \F^{\aea}_N\pa{ \pa{\f{2\ceil{y}-1}{N}}^N} \underset{\substack{\eqref{ineq:parmis}}}{\le} \;  \F^{\aea}_N\pa{  \ab{\cJ^{N,\aea,\le}_{\ceil{y} - \dl{s} \floor{\frac N2}}}  } \\
	 &\qquad \underset{\substack{\eqref{eq:J_s}}}{=} \; \ceil{y} - \floor{\tfrac N2} \dl{s} \le y +1  - \floor{\tfrac N2} \dl{s}.
 \end{align*}
We now invert the formula
\begin{align*}
	t = \pa{\f{2y -1}{N}}^N \iff y = \ud t^{\f{1}{N}} N + \f 12,
\end{align*}
to deduce that for any $t \in [1,+\infty[$, 
\begin{align*}
	\F_N^\aea(t) &\le \ud t^{\f{1}{N}} N + \f 32 - \floor{\frac N2} \dl{s} = \f N2 \pa{t^{\f 1N} + \f 3N - \dl{s} + \f {1}{N} \dl{s} \delta_{N \in 2\N +1}}  \\ 
		     & \le  \f N2 \pa{t^{\f 1N}- \dl{s} + \f 4N  },
\end{align*}
where we used that $\floor{\tfrac N2} = \f 12 \pa{N - \delta_{N \in 2\N+1}}$. In particular, the previous inequality holds for $t=k \in \Np$. 

As for obtaining a lower bound for $\F^\aea_N$, for any $y \in \R_+$ such that 
\begin{align*}
	\floor{y} \ge \max\pa{\f{N}{2e}-\f 12, \dl{s} \floor{\frac N2}},
\end{align*}
we have
\begin{align*}
&\F^{\aea}_N\pa{ \pa{\f{e(2y+1)}{N}}^N}  \ge \F^{\aea}_N\pa{ \pa{\f{e(2\floor{y}+1)}{N}}^N} \underset{\substack{\eqref{ineq:parmis}}}{\ge} \;  \F^{\aea}_N\pa{  \ab{\cJ^{N,\aea,\ge}_{\floor{y} - \dl{s} \floor{\frac N2}}}  } \\
&\qquad \underset{\substack{\eqref{eq:J_s}}}{=} \; \floor{y} - \floor{\tfrac N2} \dl{s} \ge y -1  - \floor{\tfrac N2} \dl{s}.
 \end{align*}
 Since
\begin{align*}
t = \pa{\f{e(2y +1)}{N}}^N \iff y = \ud t^{\f{1}{N}} N e^{-1} - \f 12,
\end{align*}
for any $t \in [1,+\infty[$ we have
\begin{align*}
\F_N^\aea(t) &\ge \ud t^{\f{1}{N}} N e^{-1} - \f 32 - \floor{\frac N2} \dl{s}  \ge  \f N2 \pa{t^{\f 1N} e^{-1}- \dl{s} - \f 2N  }.
\end{align*}
\end{proof}

\section{Proof of Theorem~\ref{thm:rep_can_S}}
\label{sec:proof_thm_1}

To prove Theorem~\ref{thm:rep_can_S} we will use the $\p_k^\aea$'s, defined in Section~\ref{sec:Preliminary constructions for the proofs}, to build a canonical $N$-body mixed state having a prescribed entropy $S$. Using the bound on $T(\p_k^\aea)$ will provide a bound on the kinetic energy of the mixed state, as we will see in this section.

\subsection{Building mixed states}%
\label{sub:Building mixed states}

For any vector $v$, we denote by $P_v$ the orthogonal projection onto $\C v$. For any $\theta \in \seg{0,\f{\pi}{2}}$ and any $M \in \Np$, we define the $N$-body canonical mixed state
\begin{align}\label{eq:def_mixed_state}
\gamma_N(\theta,M) &  \df \f{\pa{\cos \theta}^2}{M} \sum_{k=1}^M P_{\p_k^\aea} + \pa{\sin \theta}^2 P_{\p^\aea_{M+1}},
\end{align}
which belongs to $\cS^N\ind{can}$. Since $\ro_{\p_k^\aea} = \ro$ for any $k \in \{1,\dots,M+1\}$, $\gamma_N(\theta,M)$ has density $\ro_{\gamma_N(\theta,M)} = \ro$. The entropy of the state is
\begin{align*}
& S_{\gamma_N(\theta,M)} = - (\cos \theta)^2 \ln \f{(\cos \theta)^2}{M} - (\sin \theta)^2 \ln (\sin \theta)^2.
\end{align*}
We choose $M^* := \floor{e^S} +1$, so $S_{\gamma_N(0,M^*)} = \ln M^*  > S$ and $S_{\gamma_N\pa{\f{\pi}{2},M^*}} = 0$. Since the map $\theta \mapsto S_{\gamma_N(\theta,M^*)}$ is continuous, there is some value $\theta^* \in \seg{0,\f{\pi}{2}}$ for which $S_{\gamma_N(\theta^*,M^*)} = S$ and we define $\Gamma_N := \gamma_N\pa{\theta^*,M^*}$, respecting
\begin{align*}
\ro_{\Gamma_N} = \ro, \qquad S_{\Gamma_N} = S,
\end{align*}
and hence filling the constraints that we were targeting.

\subsection{Obtaining a bound on the kinetic energy}%
\label{sub:Obtaining a bound on kinetic}

For any $\theta \in \R$ and any $M \in \N$, we have a bound on the kinetic energy
\begin{align*}
T\pa{\gamma_N(\theta,M)} &= \f{(\cos \theta)^2}{M} \sum_{k=1}^M T\bpa{\Psi^\aea_k} + (\sin \theta)^2 T\pa{\p_{M+1}} \\
&\le \mymax{1 \le k \le M+1} T\bpa{\Psi^\aea_k}.
\end{align*}
Applying it to $\Gamma_N$, we obtain
\begin{align*}
T(\Gamma_N) & \quad \le \mymax{1 \le k \le M^*+1} T\bpa{\Psi^\aea_k}\\
&\underset{\substack{\eqref{ineq:bound_TPsi}, \eqref{eq:explicit_bound_J}}}{\le} \; \pa{1+ 4\pi^2 N^2 \pa{\pa{\floor{e^S} +2}^{\f 1N} - \dl{s} +\tfrac 4N}^2} \int_{\R^d} \ab{\na \sqrt{\ro}}^2.
\end{align*}

\subsection{Proof of~\eqref{eq:cI_can}}%
\label{sub:Proof of cI can}

Let us take $(\ro,S) \in \cI^N\ind{can}$ and $\Gamma_N \in \cS^N\ind{can}$ representing it. Then by the Hoffmann-Ostenhof inequality
\begin{align}\label{ineq:hof_ost}
\int_{\R^d} \ab{\na \sqrt{\ro_{\Gamma_N}}}^2  \le \tr \kincan\Gamma_N,
\end{align}
hence $\ro \in \cI^N$ and $\cI^N\ind{can} \subset \cI^N \times \R_+$.

Now take $(\ro,S) \in \cI^N \times \R_+$. Theorem~\ref{thm:rep_can_S} ensures the existence of a state respecting the conditions defining $\cI^N\ind{can}$, so $ \cI^N \times \R_+ \subset \cI^N\ind{can}$.

\section{Proof of Theorem~\ref{thm:rep_gc_S}}
\label{sec:proof_thm2}

We fix some normalized density $p \in \cI^1$, so $\int_{\R^d} p = 1$. We recall that from Section~\ref{sub:Harriman-Lieb orbitals}, for all $k \in \Z$ we have one-particle orbitals $\phi_k : \R^d \rightarrow \C$ such that $\ab{\phi_k}^2 = p$ and 
\begin{align}\label{ineq:kin_gc} 
\int_{\R^d} \ab{\na \phi_k}^2 \le \pa{1+ 16\pi^2 k^2} \int_{\R^d} \ab{\na \sqrt{p}}^2.
\end{align}

The proof will follow the same steps as the proof of Theorem~\ref{thm:rep_can_S}, but the pure states $\p^\aea_k$ will be replaced by grand-canonical states $\Gamma^\aea_k$.

\subsection{Constraints on ordered sets $\pa{\Gamma_k^\aea}_{k \in \N}$ of grand-canonical states}%

\subsubsection{Definition}%
\label{ssub:Definition_Gamma}

To build the grand-canonical states $\Gamma^\aea_k$, we will use the same idea as in the canonical case, but now we are allowed to populate any $n$-particle sector. We give a first solution but better fillings can be done. We start by defining
\begin{align*}
\cJ^{\aea} &:= \acs{\pa{J_n}_{1 \le n \le M} \;\Big\rvert\; M \in \N, \forall n \in \{1,\dots,M\} \; J_n \in \cJ_n^\aea \cup \{\cV_n\}},
\end{align*}
where $\cV_n$ is just a symbol for now, but we define $\Phi^\aea_{\cV_0} := 0$, being the zero of the $n$-particle sector, moreover we set $\nor{\cV_n}{\infty} := 0$.
We define the many-body pure state by imposing the constraints
\begin{itemize}
\item for any $k \in \N$, there exists $M \in \N$ and $(J_n)_{1 \le n \le M} \in \cJ^\aea$, such that
\begin{align*}
\Gamma_k^\aea = \bigoplus_{n=1}^{M} \; P_{\Phi_{J_n}^\aea},
\end{align*}
where we use the notation~\eqref{eq:notation_PhiJ}. For any $n \in \Np$ such that $J_n \neq \cV_n$, we have $\rho_{\Phi^\aea_{J_n}} = np$, hence $p_{\Gamma_k^\aea} = p$.
\item for any $\ell \in \Nz$,
\begin{align*}
\cJ^{\aea,\le}_{\ell} := \Big\{ (J_n)_{1 \le n \le 2\ell +1} \in \cJ^{\aea} \;\Big\rvert\; \mymax{1 \le n \le 2\ell +1} \nor{J_n}{\infty} \le \ell \Big\}, 
\end{align*}
we define $\Gamma_1^\aea := P_{\phi_0}$ and we impose the constraint that for any $\ell \in \Np$,
\begin{align}\label{eq:def_Gamma_k} 
\acs{\Gamma_k^\aea}_{\ab{\cJ^{\aea,\le}_{\ell-1}} +1 \le k \le \ab{\cJ^{\aea,\le}_{\ell}}} = \acs{ \bigoplus_{n=1}^{2\ell +1} \; P_{\Phi^\aea_{J_n}} \;\Bigg\rvert\;(J_n)_{1 \le n \le 2\ell +1} \in \cJ^{\aea,\le}_{\ell} \; \backslash \; \cJ^{\aea,\le}_{\ell-1}}.
\end{align}
Hence for any $n$-particle sector, for $n=1$ to $2\ell +1$, we fill it by one pure state $\Phi^\aea_{J_n}$ having density $n p$, except if $J_n = \cV_n$, in which case the $n$-particle sector is not filled.
\end{itemize}

We remark that $T(\Gamma_k^\aea)$ is not necessarily increasing in $k$.

\subsubsection{Maximum absolute value momentum}%
\label{ssub:Maximum absolute value momentum}

Let us denote by $N\ind{max}(k)$ the highest particle sector number on which $\Gamma_k^\aea$ has a non vanishing component. Let us take $k \in \N$ and consider the $k^{\text{th}}$ grand-canonical state $\Gamma_k^\aea$. There exists $(J_n)_{1 \le n \le N_{\text{max}}(k)} \in \cJ^\aea$ such that 
\begin{align*}
\Gamma_k^\aea= \bigoplus_{n=1}^{N_{\text{max}}(k)} \; P_{\Phi^\aea_{J_n}},
\end{align*}
 and we define
\begin{align*}
\F^\aea(k) \df  \mymax{a \in J_n \\ 1 \le n \le N_{\text{max}}(k)
 \\ J_n \neq \cV_n} \ab{a},
\end{align*}
so $\F^\aea(k)$ is the integer $\ell$ involved in~\eqref{eq:def_Gamma_k}. For fermions, we have that for any 
\begin{align*}
k \in \acs{\ab{\cJ^{\aea,\le}_{\ell-1}} +1 , \dots,  \ab{\cJ^{\aea,\le}_{\ell}}},
\end{align*}
$\Gamma_k^\aea$ is made up with at least one orbital having maximum absolute value momentum $\ell$, otherwise we would have $k \le \ab{\cJ^{\aea,\le}_{\ell-1}}$, hence $\F^{\text{a}}(k) = \ell$. Defining 
\begin{align*}
\F(k) := \F^{\text{a}}(k),
\end{align*}
we have
\begin{align}\label{eq:J_s_gc}
	\forall k  \in \acs{\ab{\cJ^{\aea,\le}_{\ell-1}} +1,\dots,\ab{\cJ^{\aea,\le}_{\ell}}}, \qquad  \F\pa{ k} = \ell, \qquad N\ind{max}(k) \le 2\ell +1.
\end{align}
Moreover, we can have that for any $k \in \Np$, 
\begin{align*}
\F^{\text{s}}(k) \le \F^{\text{a}}(k).
\end{align*}
 To justify this, we can argue that we could build the bosonic $\Gamma_k^\aea$ by taking the fermionic ones and replacing the antisymmetrized products by the symmetrized ones (but we have strictly more degrees of freedom to build bosonic states), even though strictly speaking this does not corresponds to the previous construction.

\subsubsection{Bound on the mean number of particles of $\Gamma_k^\aea$}%
\label{ssub:Bound on the mean number of particles }

Remark that we did not require that $\Nm_{\Gamma_k^\aea}$ increases with $k$. Nevertheless, by~\eqref{eq:J_s_gc}, for any $k \in \Np$,
\begin{align}\label{ineq:bound_nmax}
N\ind{max}(k) \le 2\F(k) +1.
\end{align}
 Moreover, in $\Gamma_k^\aea$, each $n$-particle sector is occupied by at most one pure state, hence
\begin{align}\label{ineq:bound_nmean}
\Nm_{\Gamma_k^\aea} \le \sum_{n=1}^{N\ind{max}(k)} n = \f{1}{2} N\ind{max}(k)(N\ind{max}(k)+1).
\end{align}

\subsubsection{Bound on the kinetic energy of $\Gamma_k^\aea$}%
\label{ssub:Bound on the kinetic energy of }

We can show that for any $x \in \R_+$, 
\begin{align}\label{ineq:simple_ineq} 
(x+1)(2x+1)(1+16\pi^2x^2) \le 20(2x+1)^4.
\end{align}
Note that under the previous constraints, the $\Gamma_k^\aea$'s do not necessarily have increasing kinetic energy, but we can still bound
\begin{align}\label{ineq:kin_grand_can_k}
&T(\Gamma_k^\aea) = \sum_{\substack{1 \le n \le N_{\text{max}}(k) \\ J_n \neq \cV_n}} \; \sum_{a \in J_n}  T(P_{\phi_{a}})= \sum_{\substack{1 \le n \le N_{\text{max}}(k) \\ J_n \neq \cV_n}} \; \sum_{a \in J_n}  T(\phi_{a}) \nonumber \\
&\qquad \le \Nm_{\Gamma_k^\aea} \; \mymax{1 \le j \le \F^\aea(k)} T(\phi_{j})  \nonumber \\
& \underset{\substack{\eqref{ineq:kin_gc},\eqref{ineq:bound_nmax},\eqref{ineq:bound_nmean}}}{\le} \; (\F^\aea(k)+1)(2\F^\aea(k)+1)\pa{1+ 16\pi^2 (F^\aea(k))^2} \int_{\R^d} \ab{\na \sqrt{p}}^2 \nonumber \\
& \qquad  \underset{\substack{\eqref{ineq:simple_ineq}}}{\le} \; 20 \pa{2\F^\aea(k)+1}^4 \int_{\R^d} \ab{\na \sqrt{p}}^2.
\end{align}

\subsection{Explicit bound on $\F^\aea$}%
\label{sub:Explicit bound on}
We saw in~\eqref{ineq:kin_grand_can_k} that to have a bound on the kinetic energy, we need an upper bound on $\F^\aea$. We first need the following result, which will be proved in Section~\ref{sub:Proof of LemmabJgc}.

\begin{lemma}\label{lem:bound_J_gc}
	For any $\ell \in \Np$,
\begin{align}\label{ineq:bound_J_delta_common} 
e^{\pa{\f 12 +  \dl{s}} \pa{\ell - 1}^2} \le \ab{\cJ^{\aea,\le}_\ell}.
\end{align}
\end{lemma}

From~\eqref{eq:J_s_gc} we know that $\F$ is increasing, and let us extend it from $\N$ to $[1,+\infty[$ such that it remains increasing. We do not do so for $\F^{\text{s}}$ because it is not ensured that it is increasing. For any $y \in [1,+\infty[$,
\begin{align*}
	\F\pa{  e^{\pa{\f 12 +  \dl{s}} \pa{y - 1}^2}  }  &\le \F\pa{ e^{\pa{\f 12 +  \dl{s}} \pa{\ceil{y} - 1}^2}} \underset{\substack{\eqref{ineq:bound_J_delta_common}}}{\le} \;  \F\pa{\ab{\cJ^{\aea,\le}_{\ceil{y}}}} \\
												&\underset{\substack{\eqref{eq:J_s_gc}}}{\le} \; \ceil{y} \le  y +1.
\end{align*}
Finally, since 
\begin{align*}
e^{\pa{\f 12 +  \dl{s}} \pa{y - 1}^2}= t \iff y = \pa{\f 12 + \dl{s}}^{-\f 12} \sqrt{\ln t} + 1
\end{align*}
then for any $t \in [1,+\infty[$,
\begin{align*}
\F(t) \le  \pa{\f 12 + \dl{s}}^{-\f 12} \sqrt{\ln t} + 2\le \sqrt{2} \sqrt{\ln t} + 2.
\end{align*}
We found estimates of $\F^\aea$ for both fermions and bosons to show that they are very close. Now we treat bosons and fermions with the same estimates. Since for any $k \in \Np$, $\F^\aea(k) \le \F(k)$, we deduce that for any $k \in \Np$,
\begin{align}\label{ineq:F_delta}
\F^\aea(k) \le \sqrt{2} \sqrt{\ln k} + 2.
\end{align}

\subsection{Building mixed states}%
\label{sub:Building mixed states}

We use a similar mixed state construction as in~\eqref{eq:def_mixed_state}, 
\begin{align*}
\Lambda(\theta,M) &  \df \f{\pa{\cos \theta}^2}{M} \sum_{k=1}^M \Gamma_k^\aea + \pa{\sin \theta}^2 \Gamma_{M+1}^\aea.
\end{align*}
Again, we choose $M^* := \floor{e^S} +1$, so $S_{\Lambda(0,M^*)} = \ln M^*  > S$ and $S_{\Lambda\pa{\f{\pi}{2},M^*}} = 0$. Since the map $\theta \mapsto S_{\Lambda(\theta,M^*)}$ is continuous, there is some value $\theta^* \in \seg{0,\f{\pi}{2}}$ for which $S_{\Lambda(\theta^*,M^*)} = S$ and we define $\Gamma := \Lambda\pa{\theta^*,M^*}$, respecting
\begin{align*}
\ro_{\Gamma} = \ro, \qquad S_{\Gamma} = S.
\end{align*}

\subsection{Bounds on kinetic energy and mean number of particles}%
\label{sub: on kinetic energy and mean number of particles}
First, for all $x \in \R_+$,
\begin{align}
(x+1)(2x+1) &\le 2(x+1)^2 \label{ineq:simple_1} \\
\pa{2\sqrt{2}\sqrt{\ln\pa{e^x +2}} + 5}^2 & \le 2^4 (x +4) \label{ineq:simple_2} \\  
\pa{\sqrt{2}\sqrt{\ln\pa{e^x +2}} + 3}^2 & \le 2^3 (x +4) \label{ineq:simple_3}  
\end{align}

For any $\theta \in \R$ and any $M \in \N$,
\begin{align*}
	T(\Lambda(\theta,M)) &= \f{\pa{\cos \theta}^2}{M} \sum_{k=1}^M T\pa{\Gamma_k^\aea} + \pa{\sin \theta}^2 T\pa{\Gamma_{M+1}^\aea} \le \mymax{1 \le j \le M+1} T\pa{\Gamma_{j}^\aea}
\end{align*}
so
\begin{align*}
T(\Gamma) & \le \mymax{1 \le j \le M^*+1} T\pa{\Gamma_{j}^\aea} \underset{\substack{\eqref{ineq:kin_grand_can_k},\eqref{ineq:F_delta}}}{\le} \; 20 \pa{2\sqrt{2} \sqrt{\ln (M^* +1)} + 5}^4 \int_{\R^d} \ab{\na \sqrt{p}}^2 \\
	   & \hspace{-0.5cm} \underset{\substack{M^* \le e^S +1}}{\le} \;  20 \pa{2\sqrt{2} \sqrt{\ln (e^S + 2)} + 5}^4 \int_{\R^d}\ab{\na \sqrt{p}}^2 \\
& \underset{\substack{\eqref{ineq:simple_2}}}{\le} \; 5 \times 2^{10} \pa{S + 4}^2 \int_{\R^d} \ab{\na \sqrt{p}}^2.
\end{align*}
As for the bound on the mean number of particles,
\begin{align*}
	\Nm_{\Gamma} &= \f{\pa{\cos \theta}^2}{{M^*}} \sum_{k=1}^{M^*} \Nm_{\Gamma_k^\aea} + \pa{\sin \theta}^2 \Nm_{\Gamma^\aea_{{M^*}+1}} \le \mymax{1 \le j \le M^* +1} \Nm_{\Gamma_{j}^\aea} \\
& \underset{\eqref{ineq:bound_nmax},\eqref{ineq:bound_nmean}}{\le} \; \pa{\F^\aea\pa{M^*+1} +1}\pa{2\F^\aea\pa{M^*+1} +1} \\
&\underset{\substack{\eqref{ineq:simple_1}}}{\le} \;2\pa{\F^\aea(M^*+1) +1}^2\underset{\substack{\eqref{ineq:F_delta},\eqref{ineq:simple_3}}}{\le} \;  2^4 \pa{S +4}.
\end{align*}

 \subsection{Proof of Lemma~\ref{lem:bound_J_gc} }%
 \label{sub:Proof of LemmabJgc}\text{ }

\subsubsection{An example: fermions in the grand-canonical case}
Before giving the proof, for clarity let us give an example. We define the firsts states $\Gamma_k^\text{a}$ as follows. We start by defining $\Gamma^{\text{a}}_0 := \Omega$, where $\Omega$ is the vacuum state. Each step $\ell$ will correspond to defining states having maximal absolute value momentum $\ell$, i.e. formed by $\phi_k$'s with $k\in \{-\ell,-\ell + 1, \dots, \ell\}$, filling the sectors from particle number $1$ to particle number $2\ell+1$. For the step $\ell = 0$,
\begin{align*}
	\Gamma^{\text{a}}_1 \df P_{\phi_0}.
 \end{align*}
For the step $\ell = 1$,
\begin{align*}
\Gamma^{\text{a}}_2  \df P_{\phi_1}, \qquad\qquad \Gamma^{\text{a}}_3  \df P_{\phi_{-1}},
\end{align*}
\begin{align*}
\Gamma^{\text{a}}_4 := P_{\phi_0 \wedge \phi_1},  \qquad \Gamma^{\text{a}}_5 := P_{\phi_{-1} \wedge \phi_0},  \qquad \Gamma^{\text{a}}_6  := P_{\phi_{-1} \wedge \phi_{1}},
\end{align*}
and
\begin{align*}
\begin{array}{rlrl}
\Gamma^{\text{a}}_7   &:= P_{\phi_0} \oplus  P_{\phi_0 \wedge \phi_1},      &     \Gamma^{\text{a}}_8   &:= P_{\phi_0} \oplus  P_{\phi_{-1} \wedge \phi_0}, \\
\Gamma^{\text{a}}_9   &:= P_{\phi_0} \oplus  P_{\phi_{-1} \wedge \phi_1}, &   \Gamma^{\text{a}}_{10}   &:= P_{\phi_1} \oplus  P_{\phi_0 \wedge \phi_1},  \\
\Gamma^{\text{a}}_{11}&:= P_{\phi_1} \oplus  P_{\phi_{-1} \wedge \phi_0},  &   \Gamma^{\text{a}}_{12}   &:= P_{\phi_1} \oplus  P_{\phi_{-1} \wedge \phi_1}, \\
\Gamma^{\text{a}}_{13}&:= P_{\phi_{-1}} \oplus  P_{\phi_0 \wedge \phi_1},  &   \Gamma^{\text{a}}_{14}   &:= P_{\phi_{-1}} \oplus  P_{\phi_{-1} \wedge \phi_0}, \\
\end{array}
\end{align*}
and $\Gamma^{\text{a}}_{15}:= P_{\phi_{-1}} \oplus  P_{\phi_{-1} \wedge \phi_1}, $
\begin{align*}
\Gamma^{\text{a}}_{16 + j}  := \Gamma^{\text{a}}_{j} \oplus P_{\phi_{-1} \wedge \phi_0 \wedge \phi_1}, \qquad \tx{ for } j \in \acs{0,\dots,15}.
\end{align*}

The step $\ell = 1$ is then finished, so we can deduce that $\ab{\cJ^{\text{a},\le}_1} = 31$. Then for the step $\ell = 2$, we use $\phi_2$ and continue to define the following $\Gamma^{\text{a}}_j$'s.

\subsubsection{Proof}%
\label{ssub:Proof}

There are $\ab{\cJ^{n,\aea,\le}_\ell}$ ways of building an $n$-particle state composed by orbitals having maximum absolute value momentum lower or equal to $\ell$. If we also allow choosing a no-particle state, there are $\ab{\cJ^{n,\aea,\le}_\ell} + 1$ ways of doing so. Hence filling the $n$-particles sectors from $n=1$ to $n=2\ell+1$, which is an arbitrary choice, we have, for any $\ell \ge 1$,
\begin{align*}
	\ab{\cJ^{\aea,\le}_\ell} = -1 + \prod_{n=1}^{2\ell +1} \pa{ \ab{\cJ^{n,\aea,\le}_\ell}+ 1} 
\end{align*}
 ways to build states containing maximal absolute value momentum lower or equal to $\ell$. The $-1$ summed represents the choice of taking no particle, which we do not want to consider because it does not have normalized density $p$. We can verify that this formula yields $\ab{\cJ^{\text{a},\le}_1} = 31$ as in Section~\ref{sub:Proof of LemmabJgc}. 

We define intermediate numbers $m^{\text{a}}_\ell := 2\ell +1$ for fermions and $m^{\text{s}}_\ell := 2\ell$ for bosons, i.e. $m^\aea_\ell = 2\ell + \dl{a}$. For any $\ell \ge 1$, we have the lower bound
\begin{align*}
	\ab{\cJ^{\aea,\le}_\ell}& \ge \prod_{n=1}^{2\ell +1} \ab{\cJ^{n,\aea,\le}_\ell}\ge \prod_{n=1}^{m_\ell^\aea} \ab{\cJ^{n,\aea,\le}_\ell}\underset{\substack{\eqref{eq:values_abs_J}}}{=} \; \prod_{n=1}^{m_\ell^\aea} \parmi{n}{2\ell + \dl{s} n + \dl{a}} \\
	& \underset{\substack{\eqref{ineq:parmi_fonda}}}{\ge} \; \prod_{n=1}^{m_\ell^\aea}  \pa{\f{2\ell + \dl{s} n + \dl{a}}{n} }^n,
\end{align*}
hence
\begin{align}\label{eq:intermm} 
	& \ln \ab{\cJ^{\aea,\le}_\ell} \ge \sum_{n=1}^{m_\ell^\aea} \pa{n \ln \pa{2\ell + \dl{s} n + \dl{a}}- n \ln n} \nonumber \\
& \qquad  = -\sum_{n=1}^{m_\ell^\aea} n\ln n + \dl{s} \sum_{n=1}^{2\ell} n\ln (2\ell + n)+  \dl{a} (2\ell +1) (\ell+1) \ln (2\ell +1).
\end{align}
Since $t \mapsto t \ln t$ is non-decreasing on $[1,+\infty)$ and has $\f 12 t^2 \pa{\ln t - \f 12}$ for primitive,
we compute
\begin{align}
&\int_0^b t \ln (a + t) \d t = \int_a^{a+b} \pa{t \ln t - a \ln t} \d t \\
&\qquad = \seg{\f 12 t^2 \pa{-\f 12 + \ln t}  + a \pa{t - t \ln t} }^{a+b}_a \nonumber \\
& \qquad = \f 12\seg{ t \pa{-\f t2 + t \ln t + 2a \pa{1 - \ln t}}}^{a+b}_a, \label{eq:int_interm}
\end{align}
then
\begin{align}\label{eq:int1} 
	\sum_{n=1}^{2\ell} n \ln (2\ell + n) \ge \int_0^{2\ell} t \ln (t + 2\ell) \d t
	 \underset{\substack{\eqref{eq:int_interm}}}{=} \; \ell^2 \pa{1 + 2 \ln (2\ell)}
\end{align}
and 
\begin{align}\label{eq:int3} 
\sum_{n=1}^m n \ln n \le m \ln m + \int_1^m t \ln t \, \d t = m \ln m + \ud m^2 \ln m - \f{m^2}4 + \f 14.
\end{align}

For any $x \in [1,+\infty[$, we can show that
\begin{align}\label{ineq:one_of} 
\f 32 (x-1)^2 \le 2x(x - \ln(2x)) - \f 14 
\end{align}
For bosons, $m_\ell^{\text{s}} = 2\ell$, and for any $\ell \in \Np$,
\begin{align*}
\ln \ab{\cJ^{\text{s},\le}_\ell} &\underset{\substack{\eqref{eq:intermm}}}{\ge} \; \sum_{n=1}^{2\ell} \pa{n \ln (2\ell + n) - n \ln n} \underset{\substack{\eqref{eq:int1},\eqref{eq:int3}}}{\ge} \; 2\ell \pa{\ell - \ln (2 \ell)} - \f 14 \\
& \underset{\substack{\eqref{ineq:one_of}}}{\ge} \;\f 32 (\ell-1)^2.\\
\end{align*}

For any $x \in [1,+\infty[$, we can show that
\begin{align}\label{ineq:one_of2} 
\f 12 \pa{x - 1}^2 \le x^2 + x - \pa{x + \f 12} \ln (2x +1).
\end{align}
For fermions, $m_\ell^{\text{a}} = 2\ell +1$ and we obtain, for $\ell \ge 1$,
\begin{align*}
\ln \ab{\cJ^{\text{a},\le}_\ell} & \underset{\substack{\eqref{eq:intermm}}}{\ge} \;  (2\ell +1) (\ell+1) \ln (2\ell +1) -\sum_{n=1}^{2\ell +1} n\ln n \\
&\underset{\substack{\eqref{eq:int3}}}{\ge} \; \ell^2 + \ell - \pa{\ell + \f 12} \ln (2\ell +1)  \underset{\substack{\eqref{ineq:one_of2}}}{\ge} \; \f 12 \pa{\ell - 1}^2.
\end{align*}

Finally,
\begin{align*}
\ln \ab{\cJ^{\text{a},\le}_\ell} \ge \f 12 \pa{1 + 2\dl{s}} \pa{\ell - 1}^2 .
\end{align*}

\section{Proof of Theorem \ref{thm:rep_gc_N_S}}
\label{sec:proof_thm_3_rep_gc_N_S}

\subsection{Proof of Lemma~\ref{lem:behavior_s_inv}}%
\label{sub:Proof of bound_s_inv}

For any $x \in (0,\f{1}{2})$, we compute $s'(x) = - \ln x + \ln (1-x)$ so $s'(x) \rightarrow +\infty$ when $x \rightarrow 0^+$. Moreover, $s$ is strictly increasing and strictly convex, $s(0) = 0$ and $s\pa{\f{1}{2} } = \ln 2 > \f{1}{2} $ so $x \le s(x)$ for any $x \in [0,\f 12]$.

Take $\ep > 0$. For $x \in [0,\f{1}{2} ]$, we define $h_\ep(x) := x^{\f{1}{1+\ep} }$, $g_\ep := h_\ep - s$, and we compute 
\begin{align*}
g'_\ep(x) = \f{1}{(1+\ep) x^{\f{\ep}{1+\ep} }} + \ln x - \ln(1-x)
\end{align*}
so $g_\ep'(x) \rightarrow +\infty$ as $x \rightarrow 0^+$. We have $g_\ep\pa{\f{1}{2} } = 0$ if and only if $\ep = \ep_0 := - \f{\ln 2}{\ln \ln 2} - 1 \simeq 0.89$, so when $\ep \ge \ep_0$, $g_\ep(x) \ge 0$ for any $x \in [0,\f 12]$ and when $\ep < \ep_0 $, $g_\ep(\f 12) < 0$. Moreover, the maps $s$ and $h_\ep$ are concave and as $\ep \rightarrow 0^+$, $h_\ep$ converges pointwise to $x \mapsto x$, so when $\ep < \ep_0$, there is a unique $x_\ep \in (0,\f 12)$ such that $g_\ep(x_\ep) = 0$, for $x \in [0,x_\ep]$ $g_\ep(x) \ge 0$ and for $x \in [x_\ep, \f 12]$ $g_\ep(x) \le 0$. When $\ep \rightarrow 0^+$, then $x_\ep \rightarrow 0^+$ and when $\ep \rightarrow \ep_0^-$, then $x_\ep \rightarrow \pa{\f 12}^-$, $x_{\ep_0} = \f 12$.

To sum up, when $\ep \in (0,\f 12]$, there exists $x_\ep \in (0,\f 12)$ such that for any $x \in (0,x_\ep]$,
\begin{align*}
x \le s(x) \le h_\ep(x).
\end{align*}
Since $s(x_\ep) = h_\ep(x_\ep)$, there exists $y_\ep > 0$ such that $s^{-1}(y_\ep) = h^{-1}_\ep(y_\ep)$, more precisely, the point $(y_\ep,s(y_\ep))$ is the symmetric point to $(x_\ep,s(x_\ep))$ with respect to the line of equation $x=y$, so $y_\ep = s(x_\ep)$. Moreover, $y_\ep$ decreases with $\ep$, $y_{\ep_0} = s(x_{\ep_0}) = s(\f{1}{2}) = \ln 2$ and $y_\ep \rightarrow 0^+$ as $\ep \rightarrow 0$, and for any $y \in [0,y_\ep]$, 
\begin{align*}
y^{1+\ep} = h_\ep^{-1}(y) \le s^{-1}(y) \le y.
\end{align*}

\subsection{Finding a state representing $(p,\Nm,S)$}\text{ }
\label{sub:Finding a state representing pNS}
We treat the bosonic and fermionic cases at the same time.

\subsubsection{Definition of the state}%
\label{ssub:Definition of the state}

Let us take $(p,\Nm,S) \in \cI^1 \times (\R \backslash \{0\})^2$. We recall that we denote by $\Omega$ the vacuum state of the Fock space. Let us consider $N \in \Nz$ and $M \in \Nz$ which we will fix later. We consider again the $N$-body pure states $\p_k^\aea$ defined in Section~\ref{ssub:Definition_Psi}, where the target density is $\ro := N p$. Then for any $\lambda,\alpha \in \R_+$ such that $\lambda+\alpha \le 1$, we define the grand-canonical state

\begin{align*}
\App(\lambda,\alpha) \df \lambda P_{\p_{M+1}} + \f{\alpha}{M} \sum_{k=1}^{M} P_{\p_k^\aea} + (1-\lambda-\alpha) P_{\Omega},
\end{align*}
which respects $\App(\lambda,\alpha) \in \cS\ind{gc}$. 

\subsubsection{Properties of the state}%
\label{ssub:Properties of the state}

We compute the mean number of particles
\begin{align*}
\Nm_{\App(\lambda,\alpha)} &= \lambda N + \f{\alpha}{M} \sum_{k=1}^{M} N + (1-\lambda-\alpha) \times 0 = \pa{\lambda + \alpha} N,
\end{align*}
the entropy
\begin{align*}
S_{\App(\lambda,\alpha)} &=  - \lambda \ln \lambda + \pa{\sum_{k=1}^{M} -\f{\alpha}{M}  \ln \f{\alpha}{M}  } - \pa{1-\lambda-\alpha} \ln \pa{1-\lambda-\alpha} \\
&=  - \lambda \ln \lambda +\alpha \ln M - \alpha \ln \alpha- \pa{1-\lambda-\alpha} \ln \pa{1-\lambda-\alpha},
\end{align*}
and the kinetic energy
\begin{align}\label{eq:bound_kin_rep_S_N} 
&T\pa{\App\pa{\lambda,\alpha}} = \lambda T\pa{\p_{M+1}} + \f{\alpha}{M} \sum_{k=1}^M T\bpa{\p_k^\aea} \le (\lambda + \alpha) \; \mymax{1 \le k \le M+1} T\bpa{\p_k^\aea} \nonumber \\
&\qquad \le \; \mymax{1 \le k \le M+1} T\bpa{\p_k^\aea} \underset{\substack{\eqref{ineq:bound_TPsi}}}{\le} \; \pa{1 + 16 \pi^2 \pa{F^\aea_N(M+1)}^2} \int_{\R^d} \ab{\na \sqrt{\ro} }^2 \nonumber \\
& \qquad \underset{\substack{\eqref{eq:explicit_bound_J}}}{\le} \; \pa{1+ 4\pi^2 N^2 \pa{\pa{M+1}^{\f{1}{N}} - \dl{s} + \tfrac 4N}^2 } \int_{\R^d} \ab{\na \sqrt{\ro}}^2.
\end{align}

\subsubsection{Matching the normalized density}%
\label{ssub:Matching the normalized density}

Since for any $k \in \{1,\dots,M+1\}$, $\ro_{\p_k^\aea} = \ro = N p$ and $\ro_\Omega = 0$, the normalized density is
\begin{align*}
p_{\App(\lambda,\alpha)} &= \f{\ro_{\App(\lambda,\alpha)}}{\Nm_{\App(\lambda,\alpha)}} = \f{1}{\Nm_{\App(\lambda,\alpha)}} \pa{\lambda \ro_{\p_{M+1}} + \f{\alpha}{M} \sum_{k=1}^{M} \ro_{\p_{k}} + (1-\lambda-\alpha) \ro_{\Omega} } \\
&= \f{ (\lambda + \alpha) \ro }{(\lambda + \alpha) N} = p.
\end{align*}

\subsubsection{Matching the mean number of particles}%
\label{ssub:Matching the mean number of particles}

To satisfy $\Nm_{\App(\lambda,\alpha)} = \Nm$, we impose
\begin{align*}
\lambda + \alpha = \f{\Nm}{N},
\end{align*}
and we consider now that $\alpha$ is fixed by the value of $\lambda$, and that the remaining degrees of freedom are $\lambda$, $M$ and $N$. We will impose $\Nm \le N$ to have $\lambda + \alpha \le 1$.

\subsubsection{Matching the entropy}%
\label{ssub:Matching the entropy}

For any $x \in [0,1]$ we define
\begin{align*}
\widetilde{s}(x) := -x\ln x -(1-x) \ln (1-x),
\end{align*}
which uses the same formula as $x \mapsto s(x)$ but has a larger domain, and we recall that the entropic function $s$ is defined in~\eqref{eq:def_s(x)}.

\bul We recall that $\xi \le 1$. We choose
\begin{align*}
N = \roof{\f{\Nm}{\xi(S)}}, \qquad M = \roof{e^{\pa{\f{1}{\xi(S)} + \f{1}{\Nm}}S}},
\end{align*}
so
\begin{align}\label{ineq:bound_N_rep} 
\f{\Nm}{\xi(S)} \le N \le \f{\Nm}{\xi(S)} +1, \qquad e^{S \pa{ \f{1}{\xi(S)} + \f{1}{\Nm}  }} \le M \le e^{S \pa{ \f{1}{\xi(S)} + \f{1}{\Nm}  }} + 1.
\end{align}
Hence $\f{1}{\xi(S)} + \f{1}{\Nm} \ge \f{N}{\Nm} $ so $M \ge e^{\f{N}{\Nm}S}$ and
\begin{align}\label{ineq:_on_S} 
S \le \tfrac{\Nm}{N} \ln M.
\end{align}

\bul For any $\lambda \in \seg{0,\tfrac{\Nm}{N}}$, we define
\begin{align*}
\cS\pa{\lambda} :=  S_{\App\pa{\lambda,\tfrac{\Nm}{N} -\lambda}},
\end{align*}
which is continuous. We have
\begin{align*}
\cS\pa{0} = \tfrac{\Nm}{N} \ln M + \widetilde{s}\pa{\tfrac{\Nm}{N}} \underset{\substack{\eqref{ineq:_on_S}}}{\ge} \; S, \qquad \cS\pa{\tfrac{\Nm}{N}} = \widetilde{s}\pa{\tfrac{\Nm}{N}}.
\end{align*}

\bul If $S \ge \ln 2$, since $\max_{x \in [0,1]} \widetilde{s}(x) = \ln 2 \le S$, then
\begin{align}\label{eq:cons_S}
\cS\pa{\tfrac{\Nm}{N}} \le S.
\end{align}
If $S \le \ln 2$, $\xi(S) = s^{-1}(S)$, so from~\eqref{ineq:bound_N_rep} we have $N \ge \f{\Nm}{s^{-1}\pa{S}}$, then $\f{\Nm}{N} \le s^{-1}(S) \le \f 12$ and applying $s$ yields $s\pa{\f{\Nm}{N} } \le S$. Since $\f{\Nm}{N} \le \f 12$, $s\pa{\f{\Nm}{N} } = \widetilde{s}\pa{\f{\Nm}{N} }$ and~\eqref{eq:cons_S} also holds.

In any case, i.e. for any $S \in \R_+ \backslash \{0\}$, 
\begin{align*}
\cS\pa{\tfrac{\Nm}{N}} \le S \le \cS\pa{0},
\end{align*}
and by the intermediate value theorem there exists $\lambda^* \in \seg{0,\tfrac{\Nm}{N}}$ such that $\cS(\lambda^*) = S$.

\subsubsection{Estimation of the kinetic energy}%
\label{ssub:Estimation of the kinetic energy}

The state that we use for the theorem is $\Gamma := \App\pa{\lambda^*,\f{\Nm}{N} - \lambda^* }$. Considering~\eqref{eq:bound_kin_rep_S_N}, $\ro = N p$ and $ 1 \le N$,
\begin{align*}
T\pa{\Gamma} &\le N \pa{1+ 4\pi^2 N^2 \pa{\pa{M+1}^{\f{1}{N}} + 4}^2 } \int_{\R^d} \ab{\na \sqrt{p}}^2 \\
& \le 5 \pi^2 N^3 \pa{\pa{M+1}^{\f{1}{N}} + 4}^2  \int_{\R^d} \ab{\na \sqrt{p}}^2.
\end{align*}
For any $a,b \ge 0$ and any $q \ge 1$, $q \in \N$, $\pa{a+b}^{\f 1q} \le a^{\f 1q} + b^{\f 1q}$ so since $M \ge 1$,
\begin{align*}
\pa{M+1}^{\f 1N} = \pa{M-1 + 2}^{\f 1N} \le \pa{M-1}^{\f 1N} + 2^{\f 1N} \le \pa{M-1}^{\f 1N} + 2
\end{align*}
and
\begin{align*}
T\pa{\Gamma} & \le 5\pi^2 N^3 \pa{\pa{M-1}^{\f{1}{N}} +6}^2  \int_{\R^d} \ab{\na \sqrt{p}}^2 \\
& \underset{\substack{\eqref{ineq:bound_N_rep}}}{\le} \; 5\pi^2 N^3 \pa{ e^{ \f{S}{N}  \pa{\f{1}{\Nm} + \f{1}{\xi(S)}  }} +6}^2  \int_{\R^d} \ab{\na \sqrt{p}}^2\\
& \underset{\substack{\eqref{ineq:bound_N_rep}}}{\le} \; 5\pi^2 \pa{ \f{\Nm}{\xi(S)} +1}^3 \pa{e^{S \f{\xi(S)}{\Nm} \pa{ \f{1}{\xi(S)} + \f{1}{\Nm}}}  +6}^2 \int_{\R^d} \ab{\na \sqrt{p}}^2.
\end{align*}

\subsection{Proof of~\eqref{eq:cIgc}}%
\label{sub:Proof of}

 Take
\begin{align*}
	(p,\Nm,S) \in \cI^1  \times \bbpa{\bpa{\pa{\R_+ \backslash (\N \cup \{0\}) } \times (0,+\infty)} \cup \pa{\N \times \R_+}},
\end{align*}
hence $\Nm > 0$. If $S = 0$, then $\Nm \in \N$ and we can build a $\Nm$-body pure state having density $\ro = \Nm p$ by the Harriman-Lieb construction~\cite{Harriman81,Lieb83b}. If $S > 0$, then $\Nm \notin \N$ and we use the first part of Theorem~\ref{thm:rep_gc_N_S} to obtain the existence of a appropriate state. We then have
\begin{align*}
\cI^1  \times \bbpa{\bpa{\pa{\R_+ \backslash (\N \cup \{0\}) } \times (0,+\infty)} \cup \pa{\N \times \R_+}} \subset \cI\ind{gc}.
\end{align*}

Now let us take $(p,\Nm,S) \in \cI\ind{gc}$, we have $\Gamma \in \cS\ind{gc}$ such that $p_\Gamma = p$, $S_\Gamma = S$, $\Nm_\Gamma = \Nm$ and $\tr \bbK \Gamma < +\infty$. By the Hoffmann-Ostenhof inequality 
\begin{align}\label{ineq:hof_ost_gc}
\int_{\R^d} \ab{\na \sqrt{\ro_{\Gamma}}}^2  \le \tr \bbK \Gamma
\end{align}
for grand-canonical states, we have $p \in \cI^1$. If $S = 0$, then $\Gamma$ is a pure state, hence a canonical state, and hence $\Nm \in \N$. If $S > 0$, then either $\Nm \in \N$ or $\Nm \in \R_+ \backslash (\N\cup \{0\})$, so
\begin{align*}
\cI\ind{gc} \subset \cI^1  \times \bbpa{\bpa{\pa{\R_+ \backslash (\N \cup \{0\}) } \times (0,+\infty)} \cup \pa{\N \times \R_+}}.
\end{align*}

We have hence
\begin{align*}
	\cI\ind{gc} &= \cI^1  \times \bbpa{\bpa{\pa{\R_+ \backslash (\N \cup \{0\}) } \times (0,+\infty)} \cup \pa{\N \times \R_+}} \\
&= \cI^1  \times \bbpa{\bpa{ (0,+\infty) \times (0,+\infty)} \cup \pa{\N \times \R_+}} \\
& = \cI^1  \times \bbpa{\bpa{(0,+\infty) \times \R_+} \backslash ( (\R \backslash \N) \times \{0\})}.
\end{align*}

\subsection*{Acknowledgement}
I warmly thank Mathieu Lewin for his advice. This project has received funding from the European Research Council (ERC) under the European Union's Horizon 2020 research and innovation programme (grant agreements MDFT No 725528 and EMC2 No 810367). Data sharing is not applicable to this article as no new data were created or analyzed in this study. 

\bibliographystyle{siam}
\bibliography{biblio}
\end{document}